\documentclass[twocolumn,prl]{revtex4}
\usepackage{epsfig,amssymb,amsmath}
\usepackage[english]{babel}
\usepackage{graphicx}
\usepackage{color}
\begin{document}
\title{Anomalous Gilbert Damping and Duffing Features of the SFS {\boldmath $\varphi_0$} Josephson Junction}
\author{Yu. M. Shukrinov$^{1,2}$}
\author{I. R. Rahmonov$^{1,3}$}
\author{A. Janalizadeh$^{4}$}
\author{M. R. Kolahchi$^{4}$}
\address{$^{1}$ BLTP, JINR, Dubna, Moscow Region, 141980, Russia\\
$^{2}$ Dubna State University, Dubna,  141980, Russia\\
$^{3}$ Umarov Physical Technical Institute, TAS, Dushanbe 734063, Tajikistan\\
$^{4}$ Department of Physics, Institute for Advanced Studies in Basic Sciences, P.O. Box 45137-66731, Zanjan, Iran}
\date{\today }
\begin{abstract}
We  demonstrate unusual features of phase dynamics, IV-characteristics and magnetization dynamics of the $\varphi_0$ Josephson junction at small values of spin-orbit interaction, ratio of Josephson to magnetic energy and Gilbert damping. In particular, an anomalous shift of the ferromagnetic resonance frequency with an increase of Gilbert damping is found. The ferromagnetic resonance curves show the Duffing oscillator behaviour, reflecting the nonlinear nature of Landau-Lifshitz-Gilbert (LLG) equation. Based on the numerical analysis of each term in LLG equation we obtained an approximated equation demonstrated both damping effect and Duffing oscillator features. The resulting Duffing equation incorporates the Gilbert damping in a special way across the dissipative term and the restoring force.  A resonance method for the determination of spin-orbit interaction in noncentrosymmetric materials which play the role of barrier in $\varphi_0$ junctions is proposed.
\end{abstract}
\maketitle
\paragraph*{Introduction.}
The Josephson junctions (JJ) with the current-phase relation $I=I_c \sin(\varphi -\varphi_0)$, where the phase shift $\varphi_0$  is proportional to the magnetic moment of ferromagnetic layer determined by the parameter of spin-orbit interaction, demonstrate a number of unique features important for superconducting spintronics, and modern information technology \cite{linder15,shukrinov-ufn21,bobkova20,mazanik-pra20,shukrinov-apl17,nashaat-pepan20}. The phase shift allows one to manipulate the internal magnetic moment using the Josephson current, and the reverse phenomenon which leads to the appearance of the DC component in the superconducting current \cite{buzdin-prl08,konschelle-prl09,shukrinov-prb19}.

Interactive fields can bring nonlinear phenomena of both classical, and quantum nature. A basic example is the magnons strongly interacting with microwave photons \cite{zha14}. As a result we could name Bose-Einstein condensation of such quasiparticles, i.e. magnons \cite{ser14,dem08}, and synchronization of spin torque nano-oscillators as they coherently emit microwave signals in response to d.c. current \cite{sheh05}. It is interesting that (semi)classical anharmonic effects in the magnetodynamics described by the Landau-Lifshitz-Gilbert (LLG) model in thin films or heterostructures \cite{shen20, nik20}, and the quantum anharmonicity in the cavity mangnonics \cite{elyasi20} can well be modeled by so simple a nonlinear oscillator as Duffing. The corresponding Duffing equation contains a cubic term and describes the oscillations of the various nonlinear systems \cite{moon15}.

Despite the fact that nonlinear features of LLG are studied often during a long time and in different systems, manifestation of the Duffing oscillator behavior in the framework of this equation is still not completely studied. Closer to our present investigation, in the study of the dynamics of antiferromagnetic bimeron under an alternating current, Duffing equation forms a good model, and this has applications in weak signal detection \cite{shen20, wan99, alm07}. As another application with Duffing oscillator at work, we can mention the ultra thin Co$_{20}$Fe$_{60}$B$_{20}$ layer, and its large angle magnetization precession under microwave voltage. There are also {\lq foldover\rq} features, characteristic of the Duffing spring, in the magnetization dynamics of the Co/Ni multilayer excited by a microwave current \cite{nik20, nay79, chen09}. But nonlinear features of $\varphi_0$ Josephson junctions have not been carefully studied yet.
In this Letter, we show that the Duffing oscillator helps in the understanding of the
nonlinear features of $\varphi_0$ Josephson junctions at small values of system parameters.

Coupling of superconducting current and magnetization and its manifestation in the IV-characteristics and magnetization dynamics opens the door for the resonance method determination of spin-orbit intensity in noncentrosymmetric materials playing the role of barrier in $\varphi_0$ junctions. As it is well known, the spin-orbit interaction plays an important role in modern physics, so any novel method for its determination in real materials would be very important. There are a series of recent experiments demonstrating the modification of Gilbert damping by the superconducting correlations (see Ref.\cite{silaev2020} and citations therein). In particular, the pronounced peaks in the temperature dependence of Gilbert damping  have been observed for the ferromagnetic insulator/superconductor multilayers \cite{yao2018} which might be explained by the presence of spin relaxation mechanisms like the spin-orbit scattering \cite{silaev2020}. Here, we use the noncentrosymmetric ferromagnetic material as a weak link in $\varphi_0$  junctions. The suitable candidates may be MnSi or FeGe, where the lack of inversion center comes from the crystalline structure \cite{konschelle-prl09}.

The Gilbert damping determines the magnetization dynamics in ferromagnetic materials but its origin is not well understood yet. Effect of nonlinearity on damping in the system is very important for application of these materials in fast switching spintronics devices. Our study clarifies such effects. In Ref.\cite{zhao16} the authors discuss the  experimental study of temperature-dependent Gilbert damping in permalloy (Py) thin films of varying thicknesses by ferromagnetic resonance, and provide an important
insight into the physical origin of the Gilbert damping in ultrathin magnetic films.

\begin{figure}[tph!]
\includegraphics[height=25mm]{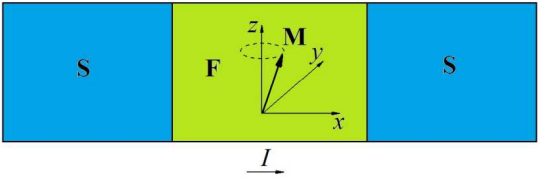}
\caption{Schematic view of SFS $\varphi_{0}$ Josephson junction. The external current applied along $x$ direction, ferromagnetic easy axis is along $z$ direction.}
\label{fig1}
\end{figure}

\begin{figure}[tph!]
\includegraphics[height=55mm]{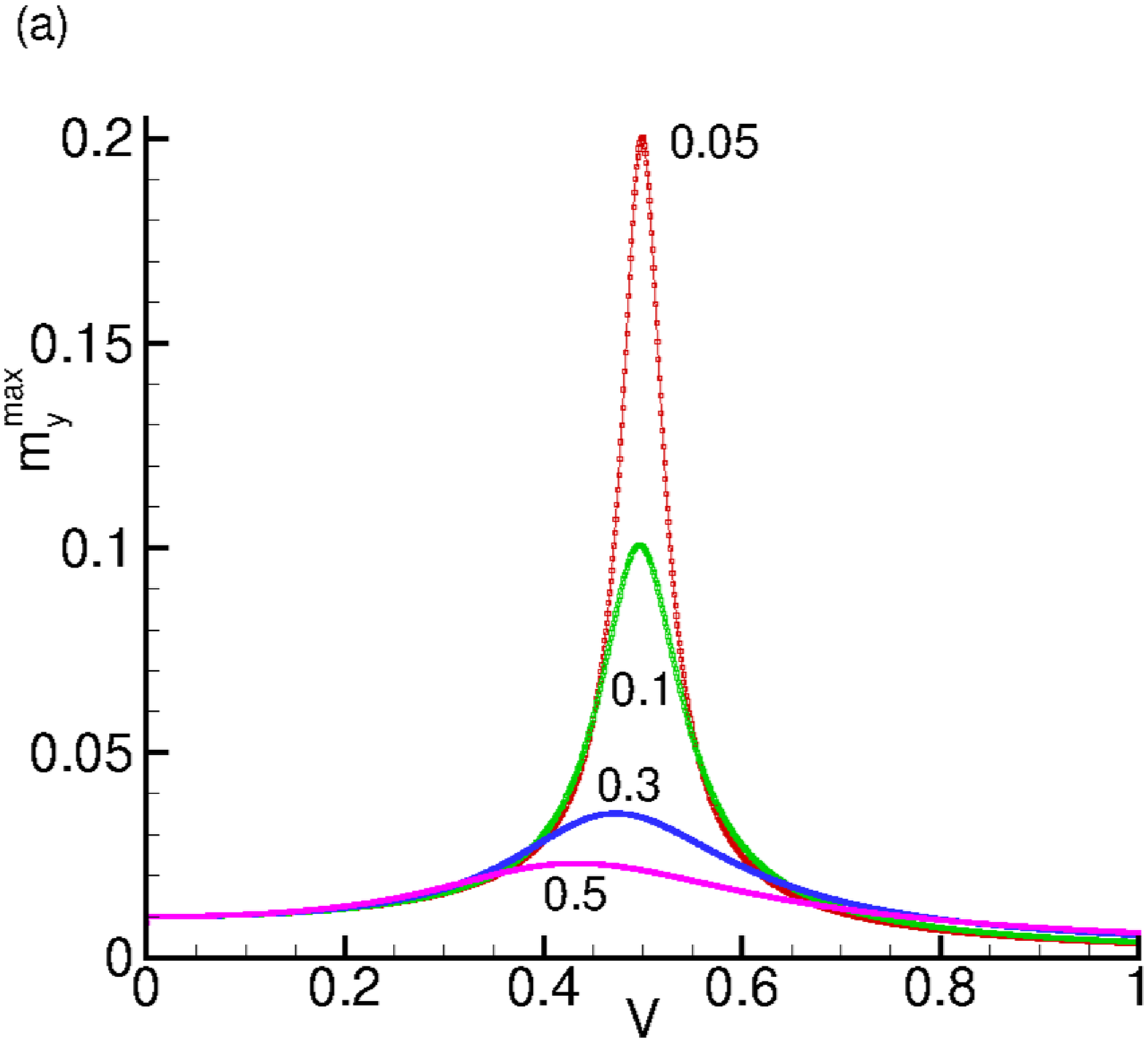}
\includegraphics[height=55mm]{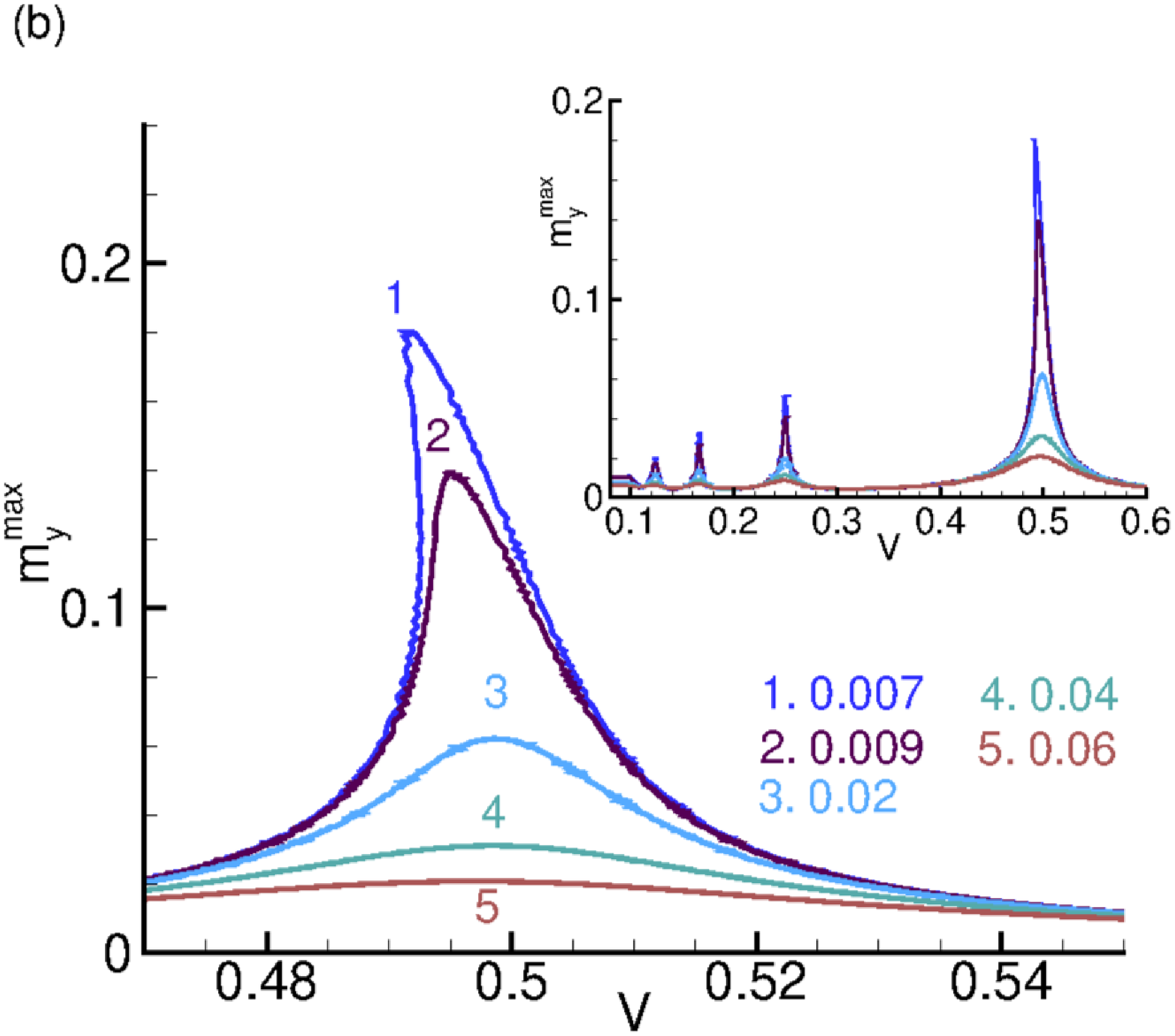}
\caption{(a) Analytical results for maximal amplitude $m_y^{max}$ in the ferromagnetic resonance region for different $\alpha$; (b) Numerical results for maximal amplitude of magnetization $m_y-$component at each values of bias current and voltage along IV-characteristics of the $\varphi_0$   junction in the ferromagnetic resonance region for various $\alpha$. Inset shows the manifestation of the resonance subharmonics. Parameters are:  $\beta_c=25$, G=0.05, r=0.05,$\omega_F=0.5$. }
\label{fig2}
\end{figure}

In this Letter we demonstrate an anomalous dependence of the ferromagnetic resonance frequency with an increase of the Gilbert damping. We find that the resonance curves demonstrate features of Duffing oscillator, reflecting the nonlinear nature of LLG equation.  The damped precession of the magnetic moment is dynamically driven by the Josephson supercurrent, and the resonance behavior is given by the dynamics of the Duffing spring. The resonance methods for the determination of spin-orbit interaction in the $\varphi_0$ junction are proposed.
\paragraph*{Model and Methods.} In the considered SFS $\varphi_{0}$ junction (see Fig.\ref{fig1}) the superconducting phase difference $\varphi$ and magnetization $\mathbf{M}$ of the F layer are two coupled dynamical variables.
Based on the LLG equation for the magnetic moment $\mathbf{M}$ with effective magnetic field ${\bf H_{eff}}$, resistively capacitively shunted junction (RCSJ) model, and Josephson relation for the phase difference $\varphi$, we describe dynamics of the SFS $\varphi_{0}$ junction by the system of equations in normalized variables
\begin{eqnarray}
\label{syseq}
\displaystyle \frac{d {\bf m}}{dt}&=&\omega_{F} {\bf h_{eff}} \times {\bf m}+\alpha\bigg({\bf m}\times \frac{d {\bf m}}{dt} \bigg),\nonumber
\vspace{0.1 cm}\\
\displaystyle {\bf h_{eff}}&=&G r\sin(\varphi - r m_{y}) {\bf\widehat{y}} + m_{z}{\bf\widehat{z}},
\vspace{0.1 cm}\\
\displaystyle \frac{dV}{dt}&=&\frac{1}{\beta_{c}}[I-V+r\frac{dm_{y}}{dt}-\sin(\varphi-r m_{y})],\nonumber
\vspace{0.1 cm}\\
\displaystyle \frac{d\varphi}{dt}&=&V,  \nonumber
\vspace{0.1 cm}
\end{eqnarray}
where ${\bf m}$ is vector of magnetization with components $m_{x,y,z}$, normalized to the $M_{0}=\|{\bf M}\|$
and and satisfying the constraint $\sum_{i=x,y,z} m_{i}^2(t)=1$, $\omega_{F}=\Omega_{F}/\omega_{c}$, $\Omega_{F}=\gamma K/\nu$ is ferromagnetic resonance frequency, $\gamma$ is the gyromagnetic ratio, $K$ is an anisotropic constant, $\nu$ is the volume of the ferromagnetic F layer, $\alpha$ is the phenomenological damping constant (Gilbert damping), ${\bf h_{eff}}$ is the vector of effective magnetic field, normalized to the $K/M_{0}$ (${\bf h_{eff}}={\bf H_{eff}}M_{0}/K$), $G = E_{J} /(K\nu)$ relation of Josephson energy to magnetic one, $r$ is a parameter of spin-orbit coupling, $\varphi$ is phase difference of JJ, $V$ is voltage normalized to the $V_{c}=I_{c}R$, $I_{c}$ critical current of JJ,  $R$ resistance of JJ, $\beta_{c} = 2eI_{c}C R^{2}/\hbar$ is McCumber parameter, $C$ is capacitance of JJ, $I$ is bias current normalized to the $I_{c}$. In this system of equation time $t$ is normalized to the $\omega_{c}^{-1}$, where $\omega_{c}=2eI_{c}R/\hbar$ is characteristic frequency. In the chosen normalization, the average voltage corresponds to the Josephson frequency $\omega_{J}$.
\paragraph*{Ferromagnetic resonance in $\varphi_0$ junction.} The ferromagnetic resonance features are demonstrated by average voltage dependence of the maximal amplitude of the $m_y$ component ($m_{y}^{max}$), taken at each value of bias current. To stress novelty and importance of our finding, we first present the analytical results for average voltage dependence of $m_{y}^{max}$ along IV-characteristics in the ferromagnetic resonance region. As it was discussed in Refs.\cite{konschelle-prl09,shukrinov-pepan20,shukrinov-ltp20},
in case $Gr\ll 1$, $m_z\approx1$, and neglecting quadratic terms $m_x$ and $ m_y $,  we get
\begin{equation}
\label{eq_sys}
\left\{\begin{array}{ll}
\displaystyle \dot{m}_{x}=\xi[-m_{y}+Gr\sin\omega_{J} t-\alpha m_{x}]
\vspace{0.2 cm}\\
\displaystyle \dot{m}_{y}=\xi[m_{x}-\alpha m_{y}],
\end{array}\right.
\end{equation}
where $\xi=\omega_{F}/(1+\alpha^{2})$. This system of equations can be written as the second order differential equation with respect to the $m_{y}$
\begin{equation}
\label{eq_d2my_linear}
\displaystyle \ddot{m}_{y}= - 2\alpha\xi\dot{m}_{y}
-\xi^{2}(1+\alpha^{2})m_{y}
+\xi^{2}Gr\sin\omega_{J} t.
\end{equation}
Corresponding solution for $m_{y}$ has the form
\begin{equation}
\label{solution2}
m_{y}(t)=\frac{\omega_{+}-\omega_{-}}{r}\sin\omega_{J} t-\frac{\alpha_{+}+\alpha_{-}}{r}\cos\omega_{J} t,
\end{equation}
where
\begin{equation}
\label{coef1}
\omega_{\pm}=\frac{Gr^{2}\omega_{F}}{2}\frac{\omega_{J}\pm\omega_{F}}{((\omega_{J}\pm\omega_{F})^{2}+(\alpha\omega_{J})^{2})},
\end{equation}
and
\begin{equation}
\label{coef2}
\alpha_{\pm}=\frac{Gr^{2}\omega_{F}}{2}\frac{\alpha\omega_{J}}{((\omega_{J}\pm\omega_{F})^{2}+(\alpha\omega_{J})^{2})}.
\end{equation}
So, $m_{y}$ demonstrates resonance with dissipation when Josephson frequency is approaching  the ferromagnetic one ($ \omega_{J} \rightarrow\omega_{F} $). The maximal amplitude $m_y^{max}$ as a function  of voltage (i.e., Josephson frequency $\omega_{J}$) at different $\alpha$, calculated using (\ref{solution2}), is presented in Fig.\ref{fig2} (a). We see the usual  characteristic variation of the resonance curve with an increase in dissipation parameter when the maximal amplitude and position of resonance pick corresponds to the damped resonance. We note that the analytical result (\ref{solution2}) were obtained in the case $Gr\ll1$.
Presented in Fig.\ref{fig2}(b) results of numerical simulations $m_y^{max}(V)$ dependence at different values of  dissipation parameter $\alpha$ demonstrate the essential differences with the results followed from the analytical consideration (\ref{solution2}). We note also that the strong coupling of the superconducting phase difference $\varphi$ and magnetization $\mathbf{M}$ of the F layer manifests itself by appearance of subharmonics of the resonance at $\omega=1/2,1/3, 1/4$ demonstrated in the inset to Fig.\ref{fig2}(b).

\begin{figure}[tph!]
\includegraphics[height=39mm]{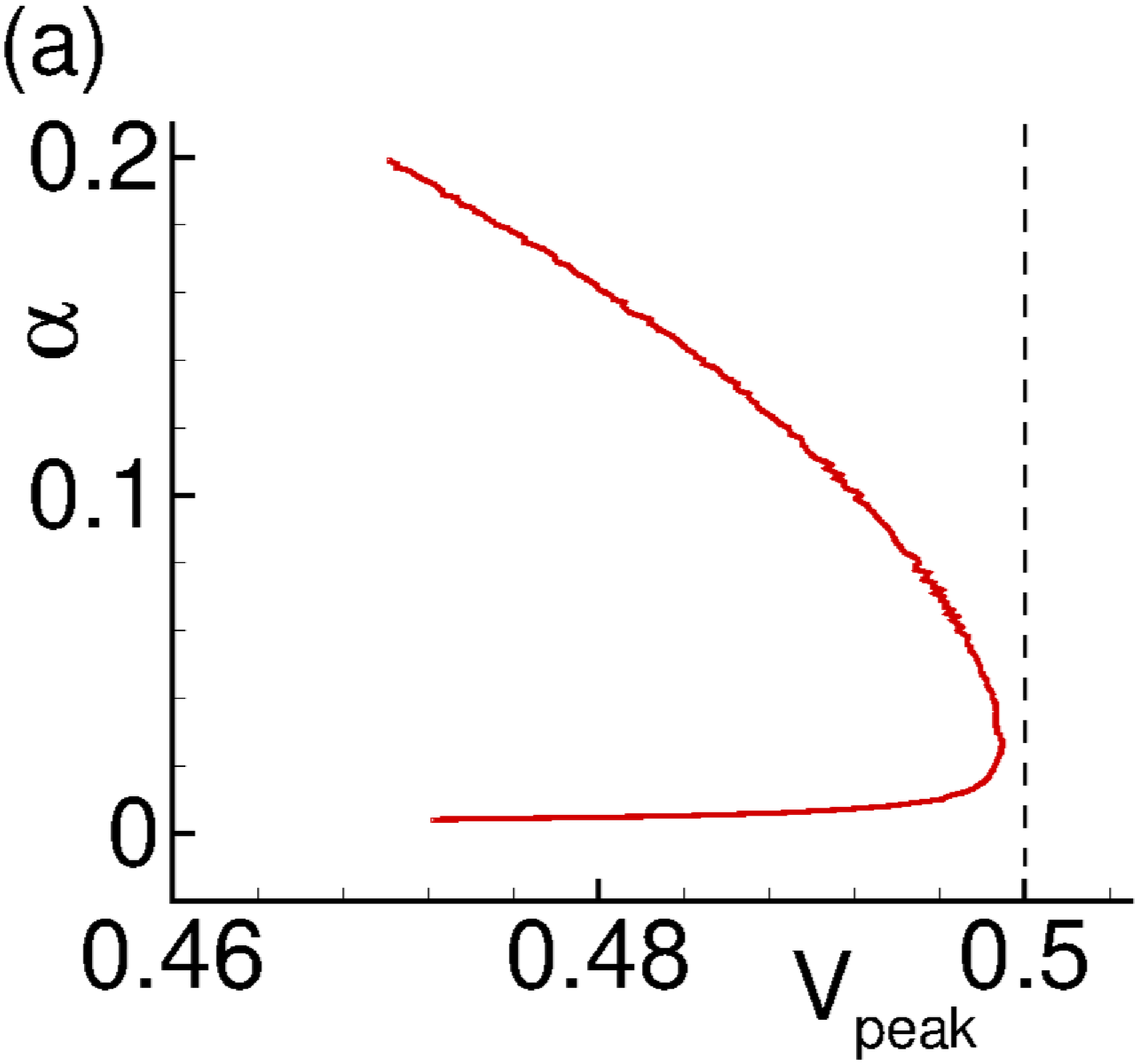}\includegraphics[height=39mm]{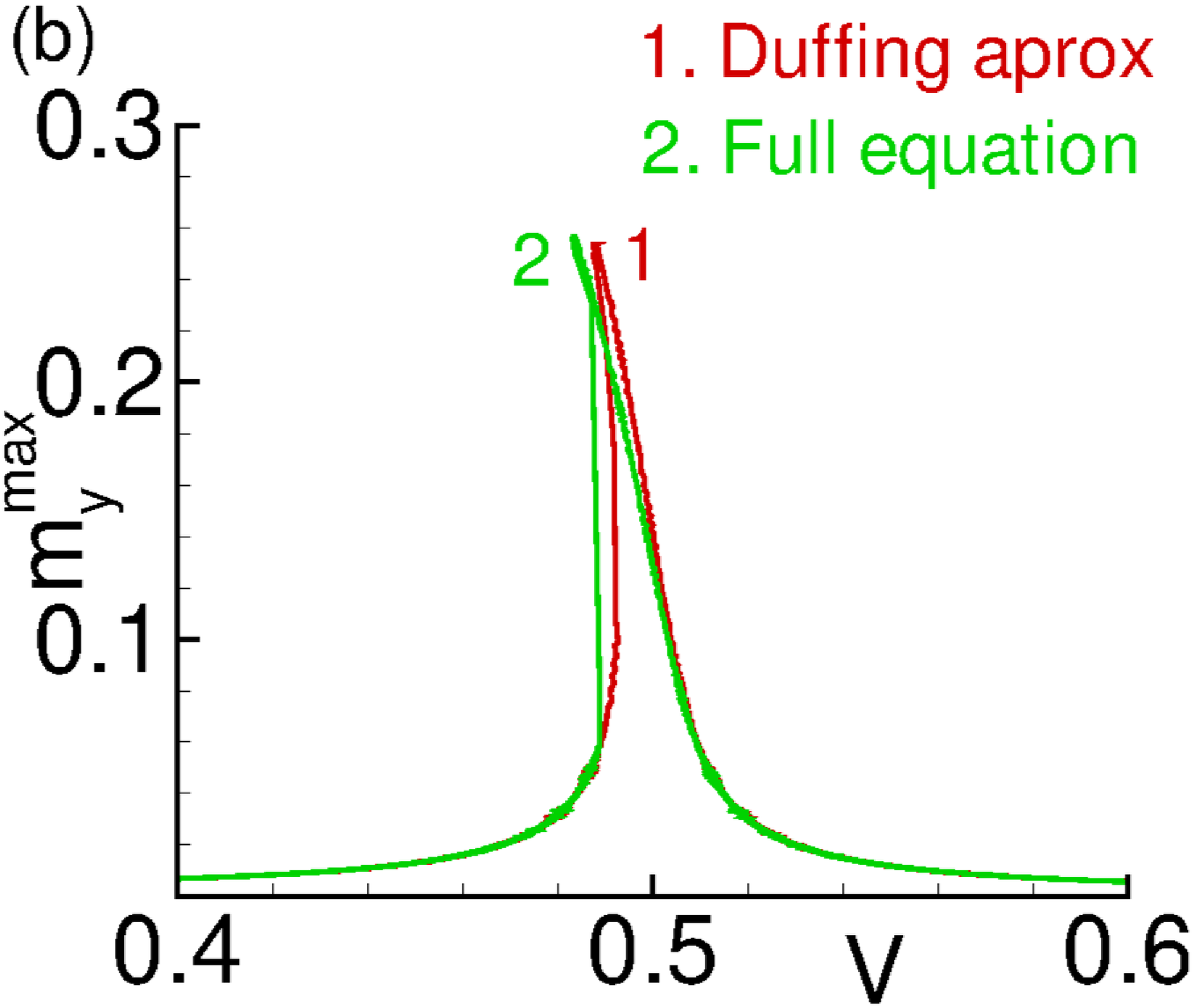}
\caption{(a) $\alpha$-dependence of the resonance curve $m_y^{max}(V)$ peak presented in Fig.\ref{fig2} in the damping parameter interval [0.006 – 0.2]. Dashed line indicates ferromagnetic resonance position; (b) Comparison of the resonance curves $m_y^{max}(V)$ calculated by full LLG equation (\ref{syseq}) and the approximate equation (\ref{eq_d2my5}).}
\label{fig3}
\end{figure}

We stress two important features followed from the presented results. First, the ferromagnetic resonance curves show the foldover effect, i.e., the features of Duffing oscillator. Different from a linear oscillator, the nonlinear Duffing demonstrates a bistability under external periodic force \cite{kovacic11}. Second,  the ferromagnetic resonance curves demonstrate an unusual dependence of the resonance frequency as a function of Gilbert damping $\alpha$.  As shown in Fig. \ref{fig3}(a), an increase in damping leads to a nonuniform change in the resonant frequency, i.e., with an increase in damping the resonance maximum shifts to $\omega_F$ at small $\alpha$, but then moves to the opposite side, demonstrating the usual damped resonance. So, with an increase in $\alpha$, unusual dependence of the resonance voltage  transforms to the usual one. For the parameters chosen, the critical value of this transformation  is around $\alpha = 0.02-0.03$. We call this unusual behaviour of the resonance maximum of $m_{y}^{max}$ as an ``$\alpha$-effect''.
Both the $\alpha-$effect and Duffing features in our system  appear due to the nonlinear features of the system dynamics at small $G,r,\alpha\ll1$. To prove it, we have carried out the numerical analysis of each term of LLG full equation (first two equations in (\ref{syseq})) for the set of model parameters $G=0.05$, $r=0.05$ $\alpha=0.005$. After neglecting the terms of order $10^{-6}$, we have
\begin{eqnarray}
\label{eq_LLG1}
\frac{\dot{m}_{x}}{\xi}&=&-m_{y}m_{z}+Grm_{z}\sin(\varphi -rm_{y})-\alpha m_{x}m_{z}^{2}, \nonumber \\
\frac{\dot{m}_{y}}{\xi}&=&m_{x}m_{z}-\alpha m_{y}m_{z}^{2},\\
\frac{\dot{m}_{z}}{\xi}&=&-Grm_{x}\sin(\varphi -rm_{y})+\alpha m_{z}(m_{x}^{2}+m_{y}^{2}), \nonumber
\end{eqnarray}
In this approximation we observe both the ``$\alpha$--effect'' and Duffing oscillator features. Neglecting here the last term $\alpha m_{z}(m_{x}^{2}+m_{y}^{2})$ in third equation for $\dot {m}_z$, which is order of $10^{-4}$, leads to the losing of the Duffing oscillator features, but still keeps alpha-effect. We note that equation (\ref{eq_LLG1}) keeps the time invariance of the magnetic moment, so that term  plays an important role for manifestation of Duffing oscillator features by LLG equation.
\paragraph*{The generalized Duffing equation for $\varphi_0$ junction.}
The LLG is a nonlinear equation and in case of simple effective field it
can be transformed to the Duffing equation \cite{moon15,shen20}. Such
transformation was used in Ref.\cite{moon15} to demonstrate the
nonlinear dynamics of the magnetic vortex state in a circular nanodisk
under a perpendicular alternating magnetic field that excites the radial
modes of the magnetic resonance. They showed Duffing-type nonlinear
resonance and built a   theoretical model corresponding to the Duffing
oscillator  from the LLG equation to explore the
physics of the magnetic vortex core polarity switching for magnetic
storage devices.

The approximated LLG system of equations (\ref{eq_LLG1}) demonstrates both $\alpha$-effect and features of Duffing oscillator. As demonstrated in the Supplemental Materials \cite{supplement}, the generalized Duffing equation for the $\varphi_0$ junction,
\begin{eqnarray}
\label{eq_d2my5}
\ddot{m}_{y}+2\xi\alpha\dot{m}_{y}+\xi^{2}(1+\alpha^{2})m_{y}\nonumber\\
-\xi^{2}(1+\alpha^{2})m_{y}^{3} &=& \xi^{2}Gr\sin\omega_{J}t.
\end{eqnarray}
can be obtained directly from the LLG system of equations.

As we see, for small enough $G$ and $r$, it is only the dimensionless damping parameter $\alpha$ in LLG that plays a role in the dynamics of the system. We can think of a harmonic spring with a constant that is hardened or softened by the nonlinear term. For a usual Duffing spring, with independent coefficients of the various terms, the resonance peak relative to the harmonic (linear) resonant frequency folds over to the smaller (softening) or larger (hardening) frequencies. In the frequency response, the interplay of the specific dependence of each coefficient on $\alpha$ plays an important role and as Fig.\ref{fig3}(a) shows, there is a particular $\alpha$ that brings the resonant frequency closest to ferromagnetic resonance.

\begin{figure}[tph!]
\includegraphics[height=60mm]{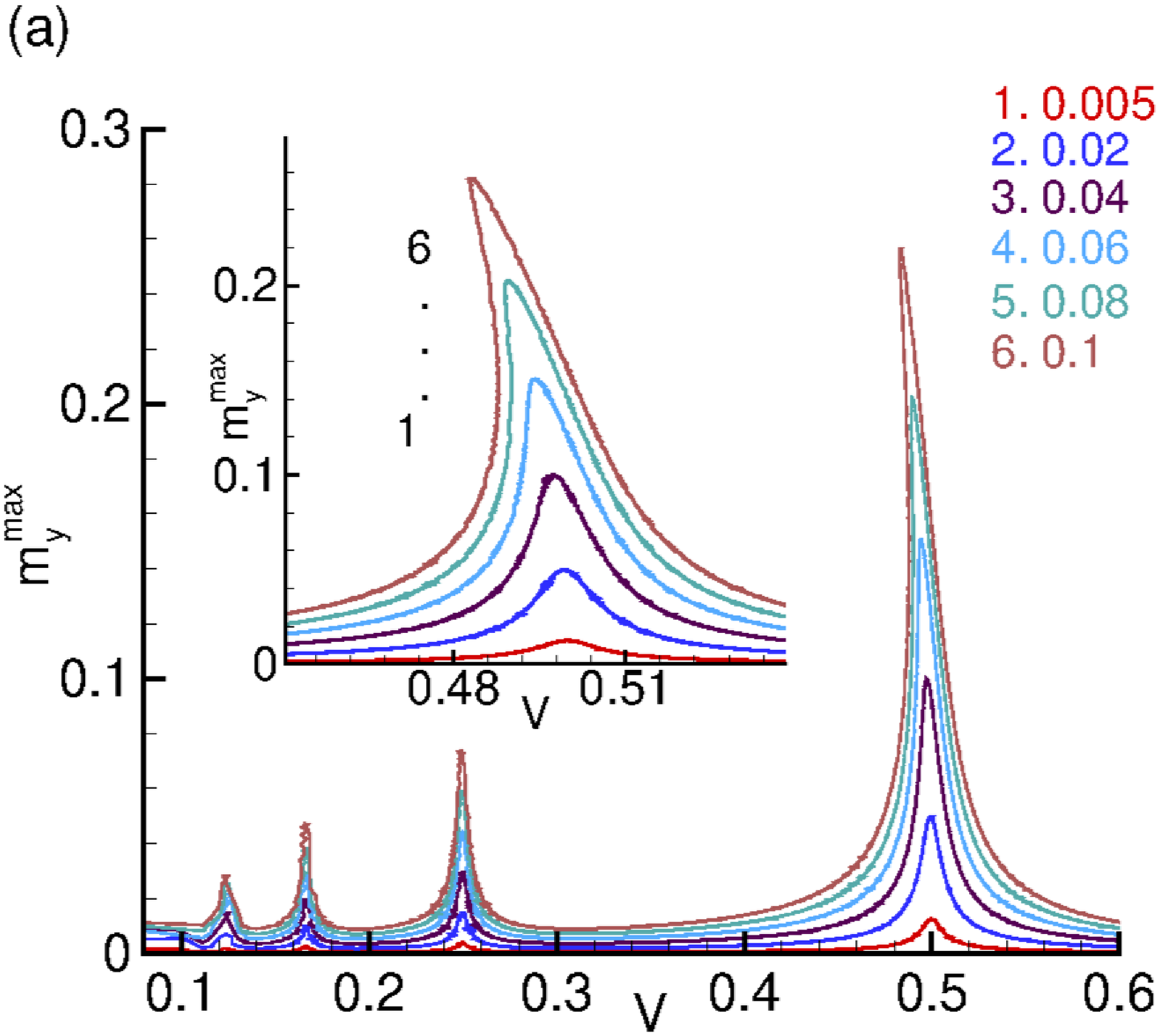}
\includegraphics[height=38mm]{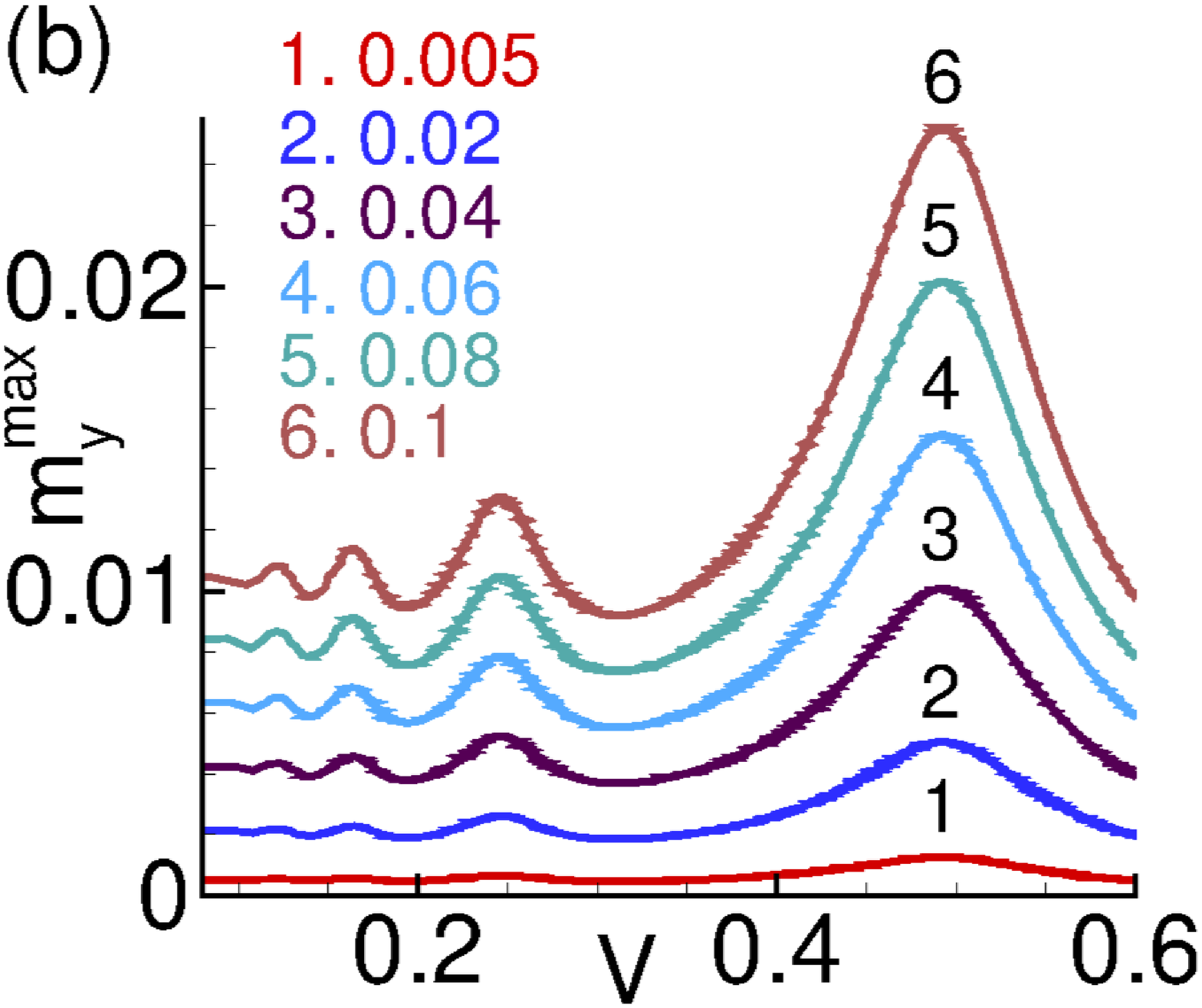}\includegraphics[height=38mm]{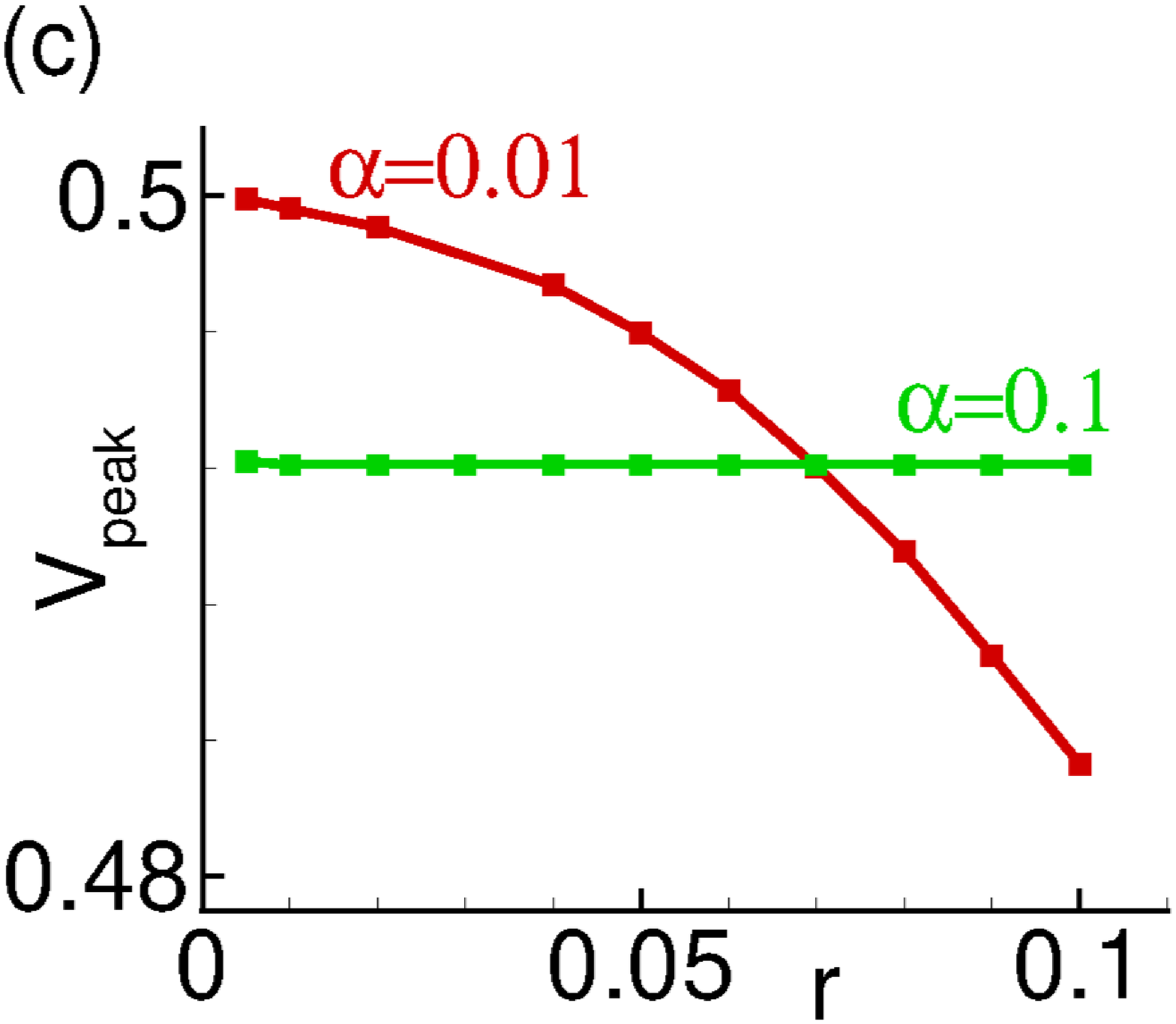}
\includegraphics[height=38mm]{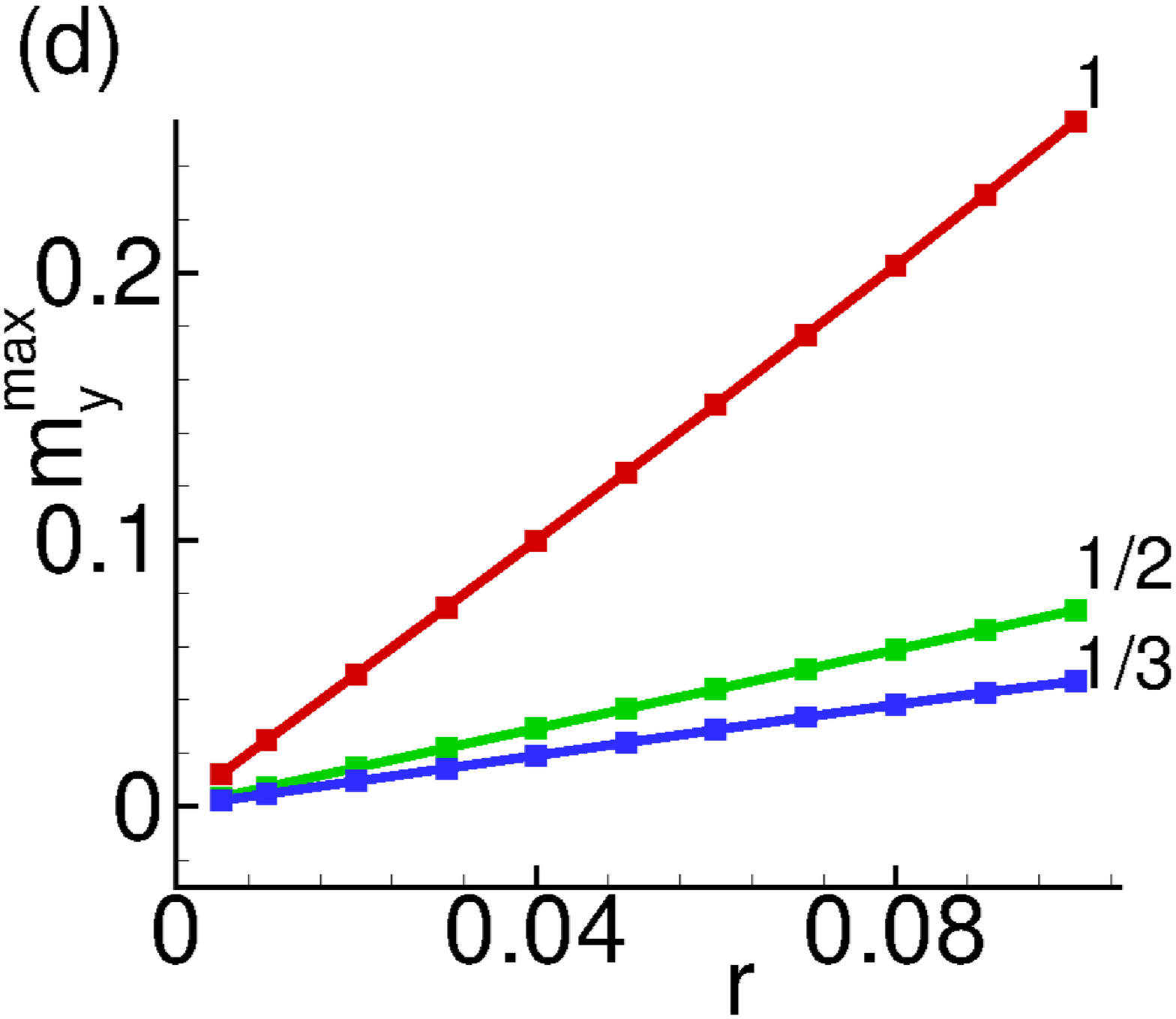}\includegraphics[height=38mm]{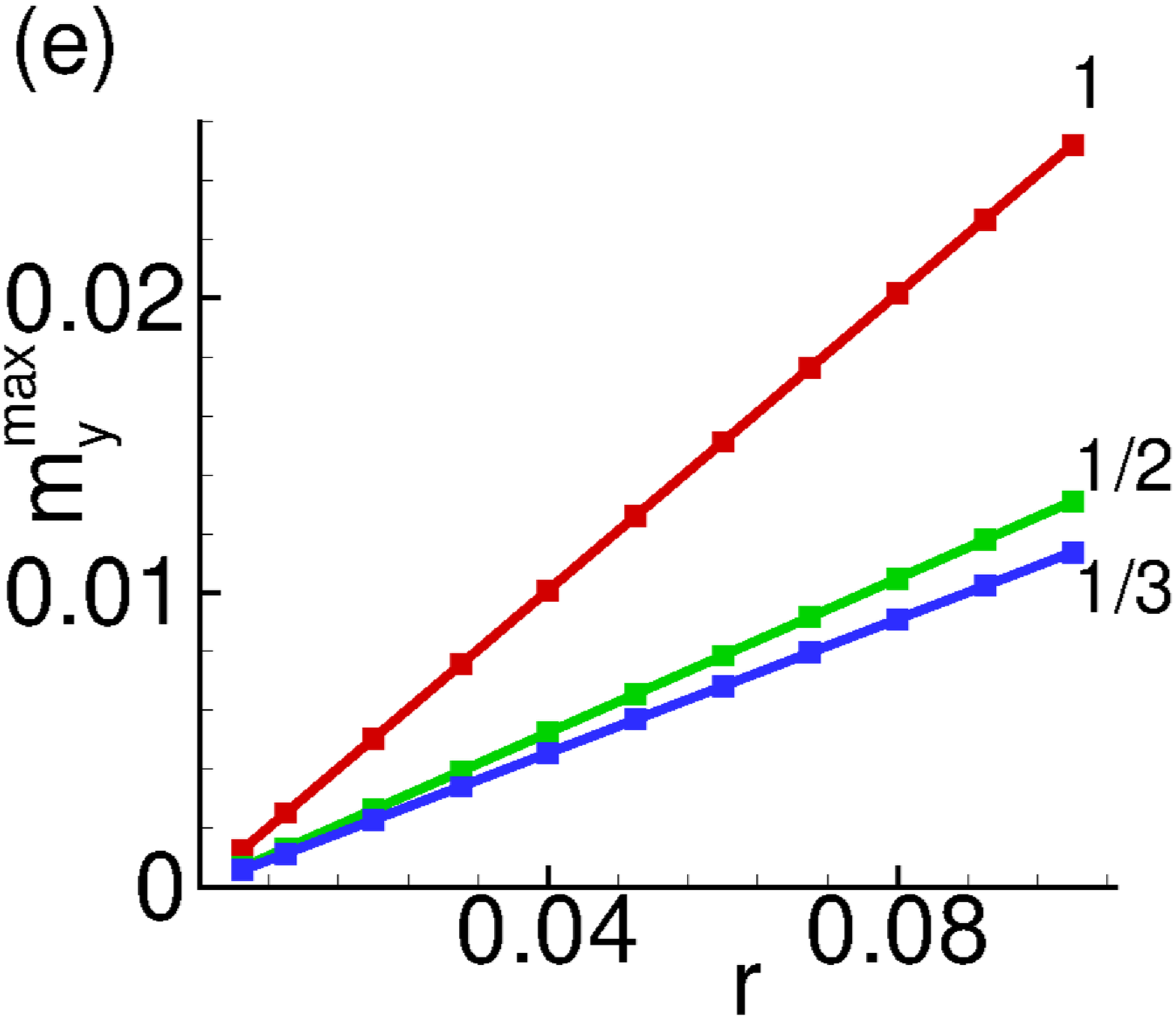}
\caption{(a) Voltage dependence of $m_y^{max}$ in the ferromagnetic resonance region at different values of spin-orbit interaction  based on (\ref{syseq}) at $G=0.05$, $\alpha=0.01$.  Inset enlarges the main harmonic; (b) The same as in  (a) for $\alpha=0.1$; (c) Shift of $m_{y}^{max}$ peak as a function of spin-orbit interaction at two values of Gilbert damping; (d) r-dependence of the main harmonic and subharmonics  peaks in case (a); (e) The same as in  (d) for the case (b). }
\label{fig4}
\end{figure}

Simulations of the $m_{y}$   dynamics in the framework of Duffing equation can explain observed foldover effect in the frequency dependence of $m_{y}^{max}$. Comparison the results followed from analytical approximate equation (\ref{eq_d2my5}) and results of full equation (\ref{syseq}) for maximal amplitude of  $m_y^{max}$ in the ferromagnetic resonance region is presented in Fig.\ref{fig3}(b). So, the magnetization dynamics in the SFS $\varphi_0$-junction due to the voltage oscillations can effectively be described by a scalar Duffing oscillator, synchronizing the precession of the magnetic moment with the Josephson oscillations.
\paragraph*{Effect of spin-orbit interactions.} As we mentioned above, the spin-orbit interaction plays an important role in different fields of modern physics. Here we have suggested a novel method for its determination  in real noncentrosymmetric ferromagnetic materials like MnSi or FeGe, where the lack of inversion center comes from the crystalline structure Ref.\cite{konschelle-prl09} and which play role a weak link in $\varphi_0$ junctions. Based on the obtained results, presented in Fig.\ref{fig4}, we propose different versions of the resonance method for the determination of spin-orbit interaction in these materials.
Particularly, in Fig.\ref{fig4}(a) we present the simulation results of maximal amplitude $m_y^{max}$ based on (\ref{syseq}) at $G=0.05$, $\alpha=0.01$  at different values of spin-orbit parameter $r$ in the ferromagnetic resonance region.  This case corresponds to the nonlinear approximation leading to the Duffing equation (\ref{eq_d2my5}). The same characteristics calculated by equation (\ref{syseq}) for larger value $\alpha=0.1$, i.e. corresponding to the linear approximation (\ref{eq_d2my_linear}) are presented in Fig.\ref{fig4}(b). As it was expected, in case $\alpha=0.01$  the foldover effect is more distinct.

In Fig.\ref{fig4}(c) the $r$-dependence of the resonance peak position, obtained from the simulation results of full equation at $\alpha=0.01$ and $\alpha=0.1$ for the same set of model and simulation parameters is demonstrated. We stress here that nonlinear features of LLG equation leading to the Duffing's shift of the $m_y^{max}$ peak of main harmonic with r presented in Fig.\ref{fig4}(c) show the manifestation of nonlinearity.

Despite the noted differences between results for $\alpha=0.01$ and $\alpha=0.1$ , we see in both cases a monotonic linear increase of $m_y^{max}$ peak of main harmonic and subharmonics with $r$ demonstrated in Fig.\ref{fig4}(d) and Fig.\ref{fig4}(e). Such linear dependence can be noted from Eq. (6) of Ref. \cite{shen20}, but the authors did not discuss it. This dependence might serve as a calibrated curve for spin-orbit interaction intensity, thus creating  the resonance methods for $r$ determination.
\paragraph*{Conclusions.} Based on the reported features of the $\varphi_0$ Josephson junction at small values of spin-orbit interaction, ratio of Josephson to magnetic energy and Gilbert damping,  we have demonstrated that the coupled superconducting current and the magnetic moments in the $\varphi_0$-junction result in the current phase relation intertwining with the ferromagnetic LLG dynamics. The ferromagnetic resonance clearly shows this interplay. In particular, an anomalous shift of the ferromagnetic resonance frequency with an increase of Gilbert damping is found. The ferromagnetic resonance curves demonstrate features of Duffing oscillator, reflecting the nonlinear nature of LLG equation. The obtained approximated equation demonstrates both damping effect and Duffing oscillator features. We have shown that due to the nonlinearity, as modeled by the generalized Duffing equation, the parameters of the system can compensate each other resulting in unusual response. The position of the maximum can shift towards and then away from the expected resonant frequency, as the damping is decreased. There are also foldover effects that was explained by the proposed model. A resonance method for the determination of spin-orbit interaction in noncentrosymmetric materials which play the role of barrier in $\varphi_0$ junctions was proposed.

The experimental testing of our results would involve SFS structures with ferromagnetic material having enough small value of Gilbert damping. Potential candidate for experimental realization could be ferromagnetic metals or insulators which have small values of damping parameter ($\alpha\sim10^{-3}-10^{-4}$). In Ref.\cite{schoen16} the authors report on a binary alloy of
cobalt and iron that exhibits a damping parameter approaching $10^{-4}$, which is comparable to
values reported only for ferrimagnetic insulators \cite{onbasli14,kelly13}. Using superconductor-ferromagnetic insulator-superconductor on a 3D topological insulator might be a way to have strong spin-orbit coupling needed for $\varphi_0$ JJ and  small Gilbert dissipation for $\alpha$-effect \cite{bobkova20}. We note in this connection that the yttrium iron garnet  YIG is especially interesting because of its small Gilbert damping ($\alpha\sim10^{-5}$).
The interaction between the Josephson current and magnetization is determined by the ratio of the Josephson to the magnetic anisotropy energy  $G =E_J/(K\nu)$  and spin-orbit interaction $r$. The value of the Rashba-type parameter $r$ in  a permalloy doped with $Pt$ ~\cite{hrabec16}  and in the ferromagnets without inversion symmetry, like MnSi or FeGe, is usually estimated to be in the range $0.1-1$. The value of the product $Gr$  in the material with weak magnetic anisotropy  $K \sim 4\times10^{-5}KA^{-3}$ \cite{rusanov04}, and a junction with a relatively high critical current  density  of $(3\times10^5 - 5\times10^6)A/cm^2$ ~\cite{robinson12} is in the range $1-100$. It gives the set of ferromagnetic layer parameters and junction geometry that make it possible to reach the values used in our numerical calculations for the possible experimental observation of the predicted effect.

Numerical simulations were funded by the project 18-71-10095 of the Russian Scientific Fund. A.J. and M.R.K. are grateful to IASBS for financial support.

\end{document}


\title{Supplemental Material to ``Anomalous Gilbert Damping and Duffing Features of the SFS {\boldmath $\varphi_0$} Josephson Junction''}

\author{Yu. M. Shukrinov$^{1,2}$}
\author{I. R. Rahmonov$^{1,3}$}
\author{A. Janalizadeh$^{4}$}
\author{M. R. Kolahchi$^{4}$}

\address{$^{1}$ BLTP, JINR, Dubna, Moscow Region, 141980, Russia\\
$^{2}$ Dubna State University, Dubna,  141980, Russia\\
$^{3}$ Umarov Physical Technical Institute, TAS, Dushanbe 734063, Tajikistan\\
$^{4}$ Department of Physics, Institute for Advanced Studies in Basic Sciences, P.O. Box 45137-66731, Zanjan, Iran}

\date{\today }

\begin{abstract}
Here, we demonstrate by numerical methods that a generalized Duffing equation can be obtained directly from LLG system of equations, for small system parameters of S/F/S junction.
\end{abstract}
\maketitle

Both the $\alpha-$effect and Duffing features obtained by LLG system of equations  appear due to the nonlinear features of its dynamics at small $G,r,\alpha\ll1$. To prove it, we have carried out the numerical analysis of each term of LLG full equation (first two equations in the equation (1) of the main text) for the set of model parameters $G=0.05$, $r=0.05$ $\alpha=0.005$. After neglecting the terms of order $10^{-6}$, we have
\begin{eqnarray}
\label{eq_LLG1}
\frac{\dot{m}_{x}}{\xi}&=&-m_{y}m_{z}+Grm_{z}\sin(\varphi -rm_{y})-\alpha m_{x}m_{z}^{2}, \nonumber \\
\frac{\dot{m}_{y}}{\xi}&=&m_{x}m_{z}-\alpha m_{y}m_{z}^{2},\\
\frac{\dot{m}_{z}}{\xi}&=&-Grm_{x}\sin(\varphi -rm_{y})+\alpha m_{z}(m_{x}^{2}+m_{y}^{2}), \nonumber
\end{eqnarray}

The procedure is as follows. Expanding $m_{z}^{n}$ in a series with the degree of $(m_{z}-1)$ we can find
\begin{eqnarray}
\label{mz_series}
m_{z}^{n}=nm_{z}-(n-1).
\end{eqnarray}
From expression $m_{x}^{2}+m_{y}^{2}+m_{z}^{2}=1$ and (\ref{mz_series}), we obtain
\begin{eqnarray}
\label{eq_mz3}
m_{z}=\frac{2-m_{y}^{2}}{2}.
\end{eqnarray}

Using approximation $\sin(\varphi -rm_{y}) = \sin(\omega_{J} t)$ in (\ref{eq_LLG1}), differentiating second equation of the system (\ref{eq_LLG1}) and substituting $\dot{m}_{x}$, $m_{x}$ and $\dot{m}_{z}$ from first second and third equations of the system (\ref{eq_LLG1}), respectively and using the expression (\ref{mz_series}), (\ref{eq_mz3}) and assuming $m_{z}=1$ only in denominators,  we come to a second order differential equation with respect to $m_{y}$

\begin{eqnarray}
\label{eq_d2my4}
\ddot{m}_{y}&=&
a_{1}\dot{m}_{y}^{3}
+a_{2}m_{y}\dot{m}_{y}^{2}
+a_{3}m_{y}^{4}\dot{m}_{y}
+a_{4}m_{y}^{2}\dot{m}_{y}
+a_{5}\dot{m}_{y}\nonumber\\
&+&a_{6} m_{y}^{5}+a_{7}m_{y}^{3}
+a_{8}m_{y}
-c_{1}\dot{m}_{y}^{2} \sin\omega_{J} t\\
&+&c_{2} m_{y}^{4}\sin\omega_{J} t
+c_{3} m_{y}^{2}\sin\omega_{J} t
+A\sin\omega_{J} t.\nonumber
\end{eqnarray}

\begin{table}[h!]
\caption{\label{tab:canonsummary}Numerical analysis of equation (\ref{eq_d2my4}) terms.}
\begin{center}
\begin{tabular}{|c|c|c|c|}
\hline
$a_{1}$ & $\frac{\alpha}{\xi}$& $a_{1}\dot{m}_{y}^{3}$& $\sim 1.76\times10^{-5}$\\
\hline
$a_{2}$ & $\alpha^{2}$& $a_{2}m_{y}\dot{m}_{y}^{2}$& $\sim 3.4\times10^{-8}$\\
\hline
$a_{3}$ & $\xi\alpha^{3}$&$a_{3}m_{y}^{4}\dot{m}_{y}$& $\sim7.7\times10^{-12}$\\
\hline
$a_{4}$ & $\xi(3\alpha-\alpha^{3})$&$a_{4}m_{y}^{2}\dot{m}_{y} $& $\sim 2\times10^{-5}$\\
\hline
$a_{5}$ & $2\xi\alpha$&$a_{5}\dot{m}_{y} $& $\sim 6\times10^{-4}$\\
\hline
$a_{6}$ & $\xi^{2}(\alpha^{2}+2\alpha^{4})$&$a_{6} m_{y}^{5} $& $\sim 5.56\times10^{-9}$\\
\hline
$a_{7}$ & $\xi^{2}(1+\alpha^{2}-\alpha^{4})$&$a_{7}m_{y}^{3}$&  $\sim  3.7 \times10^{-3}$\\
\hline
 $a_{8}$ & $\xi^{2}(1+\alpha^{2})$&$a_{8}m_{y}$& $\sim 6.1\times10^{-2}$\\
\hline
$c_{1}$ & $G r$ &$c_{1}\dot{m}_{y}^{2} \sin\varphi$& $\sim 3.6\times10^{-5}$\\
\hline
$c_{2}$ & $2\xi^{2}\alpha^{2}Gr$ &$c_{2} m_{y}^{4}\sin\varphi$&$\sim 5.3\times10^{-11}$\\
\hline
$c_{3}$ & $\xi^{2}Gr(\alpha^{2}-2)$ &$c_{3} m_{y}^{2}\sin\varphi$&$\sim 4.5\times10^{-5}$\\
\hline
$A$ & $\xi^{2}Gr$& $A\sin\omega_{J}t$ &$\sim 6.25 \times10^{-4}$\\
\hline
\end{tabular}
\end{center}
\end{table}
\vspace{0.5cm}

The numerical calculation for the used set of model parameters allows us to estimate each of the terms in the equation, as presented in Table~\ref{tab:canonsummary}.

Now, if we neglect those terms smaller than $10^{-4}$, the equation (\ref{eq_d2my4}) takes on the form of Duffing equation with damping dependent coefficients, i.e., we have a generalization of the Duffing equation
\begin{eqnarray}
\label{eq_d2my5}
\ddot{m}_{y}+2\xi\alpha\dot{m}_{y}+\xi^{2}(1+\alpha^{2})m_{y}\nonumber\\
-\xi^{2}(1+\alpha^{2})m_{y}^{3} &=& \xi^{2}Gr\sin\omega_{J}t.
\end{eqnarray}